\documentclass[twocolumn,showpacs,preprintnumbers,amsmath,amssymb]{revtex4}

\usepackage{graphicx}
\usepackage{dcolumn}
\usepackage{bm}

\begin{document}

\title{Quantum phase transition in one-dimensional Bose-Einstein condensates 
with attractive interactions}

\author{Rina Kanamoto}
\author{Hiroki Saito}
\author{Masahito Ueda}

\affiliation{Department of Physics, Tokyo Institute of Technology, Tokyo 152-8551, Japan}

\date{\today}

\begin{abstract}
Motivated by the recent development of the Feshbach technique, we studied the ground and 
low-lying excited states of attractive Bose-Einstein condensates on a one-dimensional ring 
as a function of the strength of interactions. The Gross-Pitaevskii mean-field theory predicts a 
quantum phase transition between a uniform condensate and a bright soliton, 
and a gapless singular cusp in the Bogoliubov excitation spectrum at the critical point.
However, the exact diagonalization reveals the presence of an energy gap at the critical point, 
where the singularity is smeared by quantum fluctuations. 
\end{abstract}


\maketitle
\section{Introduction}

Bose-Einstein condensates (BECs) in low dimensions~\cite{ldbec}, 
and matter-wave bright solitons in quasi-one-dimensional systems~\cite{bsoliton1,bsoliton2}
have been realized experimentally. 
In homogeneous infinite one-dimensional (1D) systems BEC does not occur 
since long-wavelength fluctuations of the phase
destroy the off-diagonal long-range order~\cite{Hoh}. 
In the presence of spatial confinement, however, BECs are possible in 1D, since the confinement introduces 
a cut-off for the long-wavelength fluctuations and hence helps maintain the long-range 
correlation of the phase.

The 1D systems of bosons with contact interactions have been solved exactly for several cases. 
For example, in the case of infinite systems, exact solutions are known for both 
repulsive~\cite{LL} and attractive interactions~\cite{MG}. 
For periodic finite systems, the exact solution is obtained for the Tonks gas 
of impenetrable bosons~\cite{G}, which corresponds to the low-density and strong repulsive-interaction limit. 
Attractive bosons confined in a finite system with periodic boundary conditions, 
however, have been studied within the mean-field theory (MFT)~\cite{Carr}, 
or based on exact solutions for a system of a few bosons~\cite{LL,ext}.
The MFT is valid when $n\xi\gg 1$, where $n$ is the particle density and $\xi$ is the healing length. 
When the strength of attractive interactions is increased, the ground state undergoes 
a transition from a uniform solution to a bright-soliton solution at a critical point. 
The purpose of this paper is to point out that even when the condition $n\xi\gg 1$ holds, the results of the MFT  
should qualitatively be modified near the critical point due to quantum fluctuations. 
We study the critical behavior of the system based on an exact diagonalization method.
Such a transition around the critical point can be studied experimentally using the Feshbach resonance, which 
can control the strength and the sign of interactions~\cite{fbMIT,fbJILA}.

The system considered in the present paper consists of $N$-bosons confined in a toroidal container 
of radius $R$ and cross section 
$S=\pi r^2$, where the condition $r\ll R$ is assumed. 
Interactions between particles at sufficiently low temperatures are well described by the contact potential
and this potential is free from divergence in 1D unlike in higher dimensions.
Low-energy scattering is characterized by the $s$-wave scattering length $a$ and 
the Hamiltonian for the system is given by
\begin{eqnarray}
\hat{{\cal H}}\!=\!\!\int_0^{2\pi}\!\!\!d\theta \left[-\hat{\psi}^{\dagger}(\theta)\frac{\partial^2}{\partial \theta^2}
\hat{\psi}(\theta)
+\frac{U}{2}\hat{\psi}^{\dagger}(\theta)\hat{\psi}^{\dagger}(\theta)\hat{\psi}(\theta)\hat{\psi}(\theta)\right],
\label{Ham}
\end{eqnarray}
where $U=\frac{8\pi aR}{S}$, $\theta$ denotes the azimuthal angle, the length and the energy are 
measured in units of $R$ and $\frac{\hbar^2}{2mR^2}$, respectively. 
The properties of the system described by the Hamiltonian~(\ref{Ham}) are characterized by 
a single, dimensionless parameter
\begin{eqnarray}
\gamma\equiv\frac{UN}{2\pi},
\end{eqnarray}
which is the ratio of the mean-field interaction energy per particle 
to the kinetic energy.
Experimentally, a quasi-1D toroidal geometry may be realized using an optical trap with 
Laguerre-Gaussian beams~\cite{toroidal}.
By tightening the confinement in the radial direction 
so that energy-level spacings in that direction exceed the interaction energy, we can restrict the atomic motion 
virtually to one dimension as described by the Hamiltonian~(\ref{Ham}). 

This paper is organized as follows. 
Section~\ref{MFT} studies stationary solutions and elementary excitations of the system within 
the Gross-Pitaevskii (GP) MFT. 
A quantum phase transition 
between uniform and bright-soliton states is found. 
The lowest excitation spectrum and the number of virtually excited particles due to interactions 
are derived analytically within the Bogoliubov approximation. 
The results indicate that the Bogoliubov theory breaks down near the critical point 
because the number of excitations diverges there. 
Section~\ref{Quantum} employs the exact diagonalization of the many-body Hamiltonian in a low-energy regime 
to examine quantum effects on the ground-state properties, low-lying spectra, and 
the position of the critical point. 
The MFT does not give any information about the two-body correlation and condensate fraction. 
In contrast, the exact diagonalization reveals that these properties deviate from those of the MFT, 
particularly in the soliton regime. 
In particular, near the critical point, the number of condensate bosons and the number of excitations 
undergo large quantum fluctuations, 
removing the singularity at the critical point and creating an energy gap.

\section{mean-field description of the quantum phase transition between uniform and soliton solutions}\label{MFT}
\subsection{Stationary solutions of the Gross-Pitaevskii equation}

\begin{figure}
\includegraphics[scale=0.5]{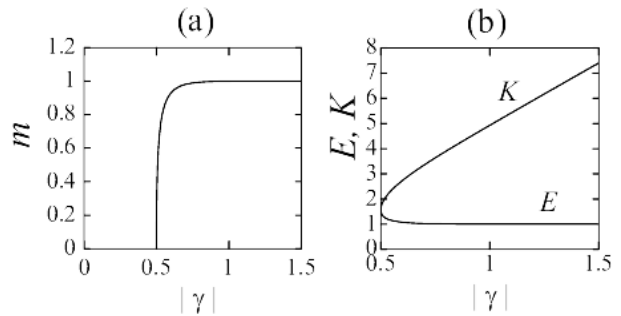}
\caption{(a) $m$ and (b) $K$ and $E$ as functions of $|\gamma|$ computed from 
Eq.~(\ref{KEeq}).}
\label{mKE}
\end{figure}

\begin{figure}
\includegraphics[scale=0.5]{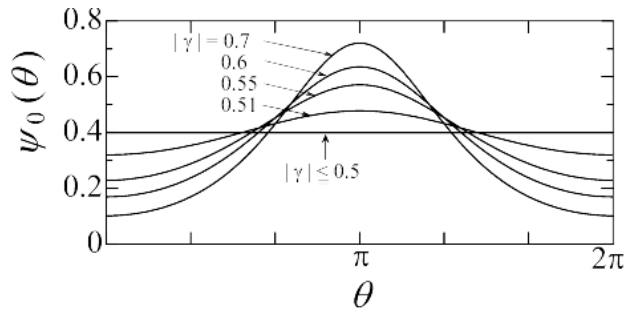}
\caption{Stationary solutions of the GP equation~(\ref{GPE}) for several values of $|\gamma|$. 
When $|\gamma|\le \frac{1}{2}$, the ground state is a uniform condensate. When $|\gamma|$ exceeds $\frac{1}{2}$, 
the ground state develops into a soliton, where $\theta_0$ in Eq.~(\ref{GPS}) is chosen to be $\pi$.}
\label{WF}
\end{figure}

A weakly interacting Bose system is well described by the Gross-Pitaevskii mean-field theory.
In a D-dimensional system with particle density $n=N/L^D$, 
``weakly interacting'' means that the healing length is much longer than the 
mean interparticle distance $n^{-1/D}$.
In 1D, this condition amounts to high density and small $s$-wave scattering length $a$. 
When these conditions are met, a condensate ``wave function'' $\psi_0$ obeys the one-dimensional GP equation
\begin{eqnarray}
i\frac{\partial \psi_0}{\partial t}=\left[-\frac{\partial^2}{\partial\theta^2}+UN|\psi_0|^2\right]\psi_0\label{GPE},
\end{eqnarray}
where $\psi_0$ is normalized as $\int_0^{2\pi} |\psi_0|^2 d\theta=1$.
A stationary solution of Eq.~(\ref{GPE}) under the periodic boundary condition 
$\psi_0(0)=\psi_0(2\pi)$ is given by \cite{Carr} 
\begin{eqnarray}
\psi_0(\theta)=\nonumber\hspace{7cm}\\
\left\{
\begin{array}{lll}
\!\!\displaystyle{\sqrt{\frac{1}{2\pi}}}&|&\!\!\!\gamma|\le |\gamma_{\rm cr}|,\\
\!\!\displaystyle{\sqrt{\frac{K(m)}{2\pi E(m)}}\ {\rm dn}\!\left(\left.\frac{K(m)}{\pi}(\theta-\theta_0)\right|m\right)}\!\!
&|&\!\!\!\gamma|>|\gamma_{\rm cr}|,
\end{array}
\right.
\label{GPS}
\end{eqnarray}
where $K(m)$ and $E(m)$ are the complete elliptic integrals of the first and the second kinds, respectively, 
and dn$(u|m)$ is a Jacobian elliptic function~\cite{math}. 
A constant phase~$\theta_0$ in Eq.~(\ref{GPS}) indicates that the soliton solution 
describes a broken symmetry state, where all states are 
degenerate with respect to $\theta_0$.
The parameter $m$ is determined from
\begin{eqnarray}
K(m)E(m)=-\frac{\pi^2\gamma}{2}\qquad (0 \le  m \le 1)\label{KEeq}.
\end{eqnarray}
It can be shown that Eq.~(\ref{KEeq}) has a solution only for $|\gamma| \geq \frac{1}{2}$, where 
the $|\gamma|$-dependences of~$m$, $K$, and $E$ are shown in Figs.~\ref{mKE}(a) and (b).
It is clear from Fig.~\ref{mKE}(a) that the mean-field critical point is given by 
$\gamma_{\rm cr}=-\frac{1}{2}$, at which 
the ground state changes from the uniform condensate to the bright soliton. 
Figure~\ref{WF} illustrates the wave functions for several interaction strength. 
We see that when $|\gamma|$ exceeds $\frac{1}{2}$, BEC begins to localize and develops 
into a soliton.

The ground-state energy per particle $\varepsilon=\varepsilon_{\rm kin}+\varepsilon_{\rm int}$, 
which is the sum of the kinetic energy $\varepsilon_{\rm kin}$ and the interaction energy $\varepsilon_{\rm int}$, 
is given for $|\gamma| \le |\gamma_{\rm cr}|$ by
\begin{eqnarray}
\varepsilon&=&\varepsilon_{\rm int}=\frac{\gamma}{2},\qquad \varepsilon_{\rm kin}=0, 
\end{eqnarray}
and for $|\gamma| > |\gamma_{\rm cr}|$ by
\begin{eqnarray}
\varepsilon\!&=&\!\frac{-K^2(m)}{3\pi^2 E(m)}\left[(2-m)E(m)+(1-m)K(m)\right],\\
\varepsilon_{\rm kin}\!&=&\!\frac{K^2(m)}{3\pi^2 E(m)}\left[(2-m)E(m)-2(1-m)K(m)\right],\\
\varepsilon_{\rm int}\!&=&\!\frac{K(m)\gamma}{6E^2(m)}\left[2(2-m)E(m)-(1-m)K(m)\right].
\end{eqnarray}
The chemical potential $\mu=\varepsilon_{\rm kin}+2\varepsilon_{\rm int}$ is therefore given by
\begin{eqnarray}
\mu=
\left\{
\begin{array}{lll}
\displaystyle{\gamma} \qquad\qquad&|&\!\!\!\gamma|\le |\gamma_{\rm cr}|,\\
\displaystyle{-\frac{K^2(m)(2-m)}{\pi^2}}\qquad\qquad&|&\!\!\!\gamma| >|\gamma_{\rm cr}|.
\end{array}
\right.
\end{eqnarray}
Figure~\ref{MFEne}(a) shows the ground-state energy $\varepsilon$, the kinetic energy $\varepsilon_{\rm kin}$, 
the interaction energy $\varepsilon_{\rm int}$ per particle, 
and the chemical potential $\mu$ as functions of $|\gamma|$.
While both $\varepsilon_{\rm kin}$ and $\varepsilon_{\rm int}$ have a cusp at $|\gamma_{\rm cr}|$, their sum 
$\varepsilon$ is everywhere smooth. 
However, the first derivative of $\varepsilon$ with respect to $|\gamma|$ has a cusp and hence its second derivative 
is discontinuous at $|\gamma_{\rm cr}|$ as shown in Fig.~\ref{MFEne}(b). 
We now study these behaviors analytically. 

\subsection{Properties of the system near the critical point and in the large-$|\gamma|$ limit}\label{MFcr}

\begin{figure}
\includegraphics[scale=0.22]{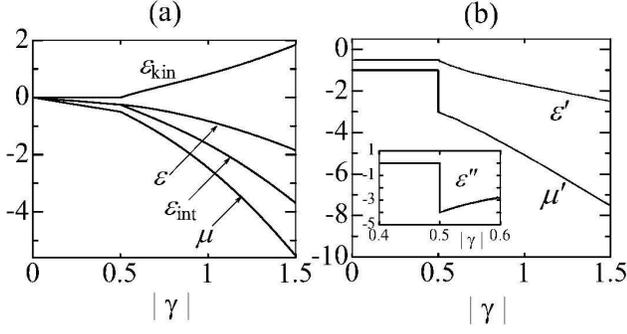}
\caption{(a) Ground-state ($\varepsilon$), kinetic ($\varepsilon_{\rm kin}$), interaction ($\varepsilon_{\rm int}$) 
energies per particle, 
and chemical potential ($\mu$). 
(b) First derivatives of the ground-state energy and chemical potential with respect to $|\gamma|$. 
Inset: Second derivative of $\varepsilon$.}
\label{MFEne}
\end{figure}

Let $\delta$ be a positive small deviation of $\gamma$ from the critical point and define 
$|\gamma_{\rm cr}|_{\pm}\equiv |\gamma_{\rm cr}|\pm\delta$. 
From Eq.~(\ref{KEeq}), we obtain $m=8\delta^{1/2}-32\delta+89\delta^{3/2}-200\delta^2+{\cal O}(\delta^{5/2})$. 
It follows then that the complete elliptic integrals can be expressed in terms of  $\delta$ as
\begin{eqnarray}
K\!\!&=&\!\!\frac{\pi}{2}\!\left(1+2\delta^{\frac{1}{2}}+\delta+\frac{1}{4}\delta^{\frac{3}{2}}
+\frac{1}{2}\delta^2\right)\!+{\cal O}(\delta^{\frac{5}{2}}),\\
E\!\!&=&\!\!\frac{\pi}{2}\!\left(1-2\delta^{\frac{1}{2}}+5\delta-\frac{33}{4}\delta^{\frac{3}{2}}
+\frac{23}{2}\delta^2\right)\!+{\cal O}(\delta^{\frac{5}{2}}).
\end{eqnarray}
In the limit of $|\gamma|\to |\gamma_{\rm cr}|_+$, the soliton solution approaches the uniform one, 
since ${\rm dn}(u|m)\to 1$ as $m\to 0$. 
In the same limit the ground-state energy per particle and 
the chemical potential behave as 
\begin{eqnarray}
\varepsilon&=& -\frac{1}{4}-\frac{\delta}{2}-2\delta^2+{\cal O}(\delta^3),\\
 \mu&=& -\frac{1}{2}-3\delta-\frac{3\delta^2}{2}+{\cal O}(\delta^3),
\end{eqnarray}
and their derivatives with respect to $|\gamma|$ are therefore given by
\begin{eqnarray}
\varepsilon'= -\frac{1}{2}-4\delta+{\cal O}(\delta^2),\quad 
\mu'= -3-3\delta+{\cal O}(\delta^2),\nonumber\\
\varepsilon_{\rm kin}'= 2-5\delta+{\cal O}(\delta^2),\quad 
\varepsilon_{\rm int}'= -\frac{5}{2}+\delta+{\cal O}(\delta^2).\label{limit1}
\end{eqnarray}
In the limit of $|\gamma|\to |\gamma_{\rm cr}|_-$, however, we obtain 
\begin{eqnarray}
\varepsilon'=\varepsilon_{\rm int}'= -\frac{1}{2},\qquad 
\varepsilon_{\rm kin}'= 0,\qquad \mu'= -1.\label{limit2}
\end{eqnarray}
Equations~(\ref{limit1}) and (\ref{limit2}) confirm that both $\varepsilon'_{\rm kin}$ and 
$\varepsilon'_{\rm int}$ are discontinuous at $|\gamma_{\rm cr}|$ as 
$\varepsilon_{\rm int}'(|\gamma_{\rm cr}|_+)-\varepsilon_{\rm int}'(|\gamma_{\rm cr}|_-)=-2$, 
$\varepsilon_{\rm kin}'(|\gamma_{\rm cr}|_+)-\varepsilon_{\rm kin}'(|\gamma_{\rm cr}|_-)=2$, 
$\varepsilon'$ is continuous everywhere. 
However, the second derivative of the ground-state energy jumps at the critical point by an amount 
\begin{eqnarray}
\varepsilon ''(|\gamma_{\rm cr}|_+)-\varepsilon ''(|\gamma_{\rm cr}|_-)=-4.
\end{eqnarray}
The first derivative of the chemical potential $\mu'$ is also discontinuous at $|\gamma_{\rm cr}|$ by an amount 
\begin{eqnarray}
\mu'(|\gamma_{\rm cr}|_+)-\mu'(|\gamma_{\rm cr}|_-)=-2.
\end{eqnarray}
The derivatives of the ground-state energy and the chemical potential are shown in Fig.~\ref{MFEne}(b).

The compressibility $\kappa$ is related to the first derivative of the chemical potential and 
the second derivative of the ground-state energy with respect to $|\gamma|$ as
\begin{eqnarray}
\frac{1}{\kappa}=n^2\frac{d\mu}{dn}=n|\gamma|\frac{d\mu}{d|\gamma|}
=\frac{|\gamma|^2}{2\pi N}\frac{d^2\varepsilon}{d|\gamma|^2}.
\end{eqnarray}
Since $\mu'$ and $\varepsilon''$ are discontinuous at $|\gamma_{\rm cr}|$, 
this point may be regarded as the critical point of a quantum phase transition between the uniform 
and soliton states at $T=0$.

In the strong-interaction regime $|\gamma|\gg |\gamma_{\rm cr}|$,
Fig.~\ref{mKE} shows that $m$ is almost equal to $1$ and that the complete elliptic integrals are approximated as 
$E\cong 1$ and $K\cong\frac{\pi^2}{2}|\gamma|$.
The soliton solution then reduces to the solution of the GP equation in infinite space~\cite{ZS}, 
\begin{eqnarray}
\psi_0(\theta)\cong\sqrt{\frac{\pi|\gamma|}{4}}\ {\rm sech}\!\!\left[\frac{\pi^2|\gamma|}{8}(\theta-\theta_0)\right].
\end{eqnarray}
In this limit, the ground-state energy and chemical potential are given in terms of $|\gamma|$ as
\begin{eqnarray}
\varepsilon\cong -\frac{\pi^2}{12}|\gamma|^2,\qquad \mu\cong -\frac{\pi^2}{4}|\gamma|^2.
\end{eqnarray}

\subsection{Low-lying Excitations}

Low-lying excitations of the condensate can be obtained by solving the Bogoliubov - de Gennes equations
\begin{eqnarray}
({\cal L}+2UN|\psi_0|^2)u_n+UN\psi_0^2 v_n&=&\lambda_n u_n,\nonumber\\
({\cal L}+2UN|\psi_0|^2)v_n+UN\psi_0^{\ast 2} u_n&=&-\lambda_n v_n,\label{BdG}
\end{eqnarray}
where ${\cal L}\equiv -\frac{\partial^2}{\partial\theta^2}-\mu$, and $\psi_0$ is 
the ground-state solution of the GP equation.

Associated with the $U(1)$ symmetry breaking in the soliton state, is a Goldstone mode 
that translates the soliton along the torus without changing its shape:
\begin{eqnarray}
u_0(\theta)\!\!\!&=&\!\!\!v_0(\theta)\nonumber\\
=\!\!\!&{\rm sn}&\!\!\!\!\left(\!\!\!\left.\frac{K(m)}{\pi}(\theta-\theta_0)\right|\!m\!\!\right)\!
{\rm cn}\!\left(\!\!\!\left.\frac{K(m)}{\pi}(\theta-\theta_0)\right|\!m\!\!\right),\\
\lambda_0&=&0.
\end{eqnarray}
Since $u_0$ and $v_0$ are proportional to $\partial\psi_0(\theta)/\partial\theta$, 
the excitation of this mode translates $\psi_0(\theta)$ 
to $\psi_0(\theta+\delta\theta)$.

Excitation energies from the uniform ground state $\psi_0=1/\sqrt{2\pi}$ are easily found to be
\begin{eqnarray}
\lambda_l=\sqrt{l^2(l^2+2\gamma)}\label{Bh}\ ,\qquad l=\pm 1, \pm 2,\dots,\label{BexU}
\end{eqnarray}
where $l$'s are the angular momenta of the excitations.
When $\gamma<-\frac{1}{2}$, the first excitation energies corresponding to $l=\pm 1$ become pure imaginaries, 
and therefore the uniform state becomes dynamically unstable. 
This implies an appearance of a new ground states that is a bright soliton state. 

\begin{figure}[h]
\includegraphics[scale=0.27]{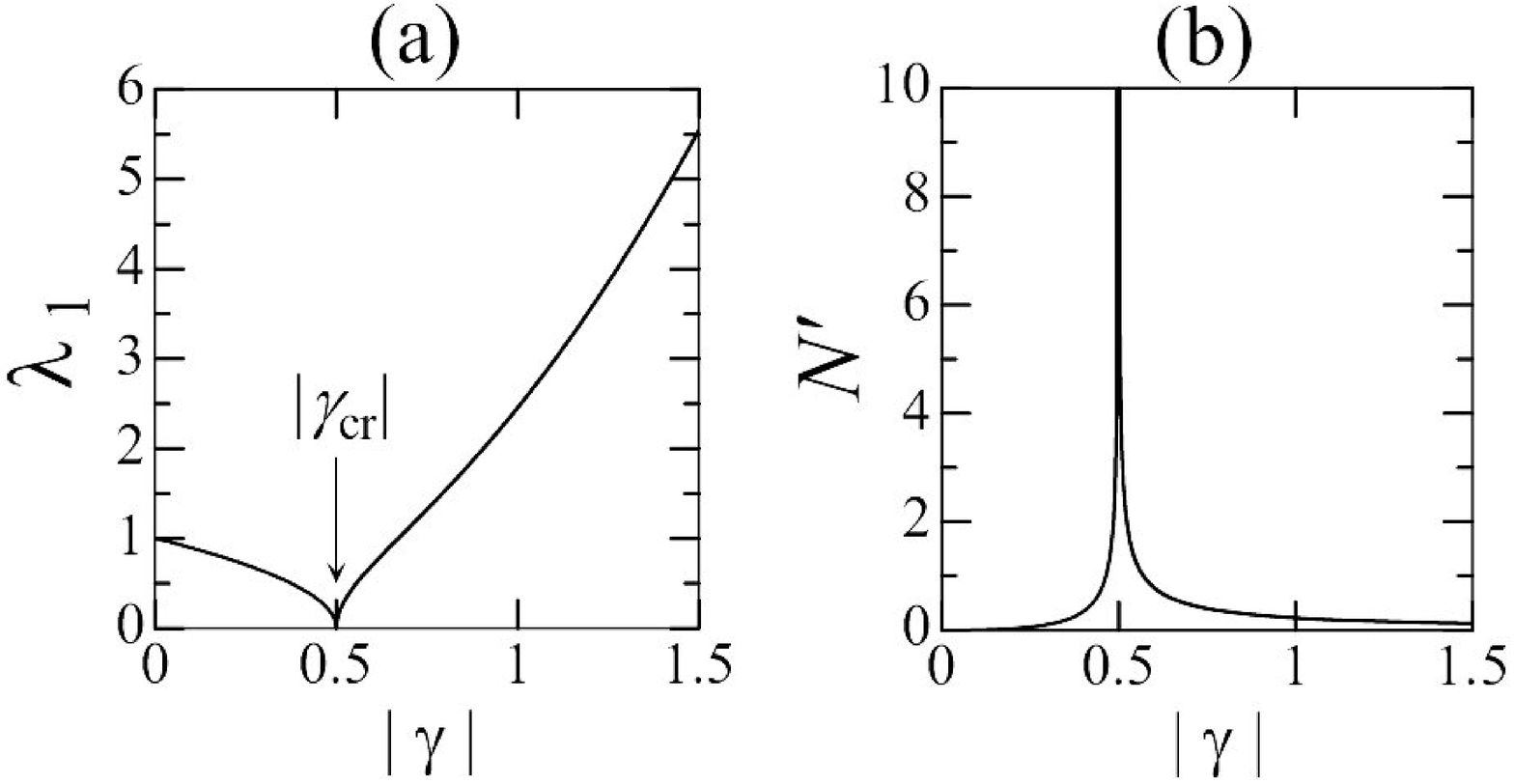}
\caption{(a) First excitation energy $\lambda_1$ and (b) the number of virtually excited particles $N'$ 
as functions of $|\gamma|$ obtained with the Bogoliubov approximation. $\lambda_1$ has a cusp and $N'$ 
diverges at $|\gamma|=0.5$.}
\label{BogEX}
\end{figure}

Let us now consider the excitation from the soliton. 
The lowest (nonzero) eigenvalue $\lambda_1$ and the corresponding amplitudes $u_1,v_1$ can be obtained analytically as
\begin{eqnarray}
\lambda_1&=&\frac{K^2(m)m}{\pi^2},\label{Bs}\\
u_1(\theta)&=&{\cal N}_1{\rm sn}^2\left(\left.\frac{K(m)}{\pi}(\theta-\theta_0)\right|m\right),\\
v_1(\theta)&=&-{\cal N}_1{\rm cn}^2\left(\left.\frac{K(m)}{\pi}(\theta-\theta_0)\right|m\right)\label{BV},
\end{eqnarray}
where the normalization constant ${\cal N}_1$ is determined from 
$\int_0^{2\pi}d\theta(|u_1(\theta)|^2-|v_1(\theta)|^2)=1$ as 
\begin{eqnarray}
{\cal N}_1^2=\frac{mK(m)}{2\pi\left[(2-m)K(m)-2E(m)\right]}.
\end{eqnarray}
Near the critical point 
$|\gamma|\to|\gamma_{\rm cr}|_+$ and in the strong-interaction limit $|\gamma|~\gg~|\gamma_{\rm cr}|$, 
Eq.~(\ref{Bs}) is simplified as 
\begin{eqnarray}
\lambda_1\cong
\left\{
\begin{array}{lll}
\displaystyle{2\sqrt{\delta}}&,&\qquad |\gamma|\to |\gamma_{\rm cr}|_+,\\
\displaystyle{\frac{\pi^2|\gamma|^2}{4}}&,&\qquad |\gamma|\gg |\gamma_{\rm cr}|\label{BexS}.
\end{array}
\right.
\end{eqnarray}
From Eqs.~(\ref{BexU}) and (\ref{BexS}), we find that the Bogoliubov spectrum possesses the same critical exponent 
on either side of $\gamma_{\rm cr}$ as

\begin{eqnarray}
\lambda_1=
\left\{
\begin{array}{lll}
\displaystyle{\sqrt{2}|\gamma-\gamma_{\rm cr}|^{\frac{1}{2}}}&,&\qquad |\gamma|\to|\gamma_{\rm cr}|_{-}\\
\displaystyle{\ 2\ |\gamma-\gamma_{\rm cr}|^{\frac{1}{2}}}&,&\qquad |\gamma|\to|\gamma_{\rm cr}|_{+}.
\end{array}
\right.
\end{eqnarray}
At the critical point, therefore, the spectrum is gapless and exhibits a cusp as shown in Fig.~\ref{BogEX}(a).

The number of virtually excited particles $N'$ is given by 
\begin{eqnarray}
N'=\int_0^{2\pi}\sum_{n\ne 0}|v_n(\theta)|^2d\theta=N-N_0, 
\end{eqnarray}
where $N_0$ is the number of condensate particles. 
Since the quantum depletion is attributed mainly to the particles with the lowest excitation energies 
in the weak interaction regime, $N'$ is obtained from the hole amplitudes $v_1$ as 
\begin{eqnarray}
N'\cong
\left\{
\begin{array}{lll}
\displaystyle{\frac{1+\gamma}{\sqrt{1+2\gamma}}-1} \qquad\qquad\qquad\qquad |\gamma|<|\gamma_{\rm cr}|,&\ &\\
\displaystyle{\frac{(1-m)(2-3m)K(m)+2(2m-1)E(m)}{3m\left[(2-m)K(m)-2E(m)\right]}} &\ &\\
\qquad\qquad\qquad\qquad\qquad\qquad\qquad |\gamma|>|\gamma_{\rm cr}|. \label{depletion}&\ &
\end{array}\label{depB}
\right.
\end{eqnarray}
The result for $|\gamma|<|\gamma_{\rm cr}|$ represents the number of virtually excited particles 
in the uniform state, which behaves as $\delta^{-\frac{1}{2}}$ in the limit of $|\gamma|\to |\gamma_{\rm cr}|_{-}$, 
while the result for $|\gamma|>|\gamma_{\rm cr}|$ represents that in the soliton state, 
which behaves as $\delta^{-3/2}$ in the limit of 
$|\gamma|\to |\gamma_{\rm cr}|_+$; 
both of them diverge at $\gamma_{\rm cr}=-\frac{1}{2}$ as shown in Fig.~\ref{BogEX}(b).
This result clearly shows the breakdown of the Bogoliubov approximation near the critical point. 
To examine the properties of the system near the critical point, 
we therefore need to go beyond the Bogoliubov approximation.

\section{Many-body effects near the critical point}\label{Quantum}

The excitation energy vanishes and the number of virtually excited particles diverges at $|\gamma_{\rm cr}|$ 
according to the Bogoliubov theory. Quantum fluctuations are thus expected to play a crucial role 
in determining the properties of the system near the critical point. 
Expanding the field operator $\hat{\psi}$ in terms of plane waves as
\begin{eqnarray}
\hat{\psi}(\theta,t)=\frac{1}{\sqrt{2\pi}}\sum_l \hat{c}_l(t)e^{il\theta},
\end{eqnarray}
where $\hat{c}_l$ is the annihilation operator of a boson with angular momentum $l$, 
we rewrite the Hamiltonian~(\ref{Ham}) in the second quantized form as
\begin{eqnarray}
\hat{{\cal H}}=\sum_{l}l^2\hat{c}_l^{\dagger}\hat{c}_l+\frac{U}{4\pi}\sum_{klmn}\hat{c}^{\dagger}_k
\hat{c}_l^{\dagger}\hat{c}_m\hat{c}_n\delta_{m+n-k-l}\label{Hamiltonian1}.
\end{eqnarray}

We diagonalize the Hamiltonian~(\ref{Hamiltonian1}) exactly within the Hilbert subspace spanned by the bases 
$|n_{-L},\dots, n_{-1},n_0,n_1,\dots,n_{L}\rangle$ 
subject to conservations of the total number of particles and the total angular momentum
\begin{eqnarray}
\sum_{l=-L}^L n_l=N, \qquad \sum_{l=-L}^L ln_l=0, \label{CC}
\end{eqnarray}
where $n_l$ denotes the number of bosons occupying the state with angular momentum $l$, and 
$L$ is the cut-off of the angular momentum.

As will be confirmed below, the inclusion of terms with $|l|\le 2$ states provides sufficient accuracy 
for our purposes, since we are interested in the properties of the system near the critical point. 
Although higher momentum states $|l|\ge 3$ become significant in the strong-interaction regime 
($|\gamma|\gg |\gamma_{\rm cr}|$), the results obtained by the truncation of the $|l|\ge 3$ states 
are quantitatively correct. 
We will therefore restrict ourselves to the states with $|l|\le 2$ unless otherwise stated.

\subsection{Crossover between the uniform and the soliton ground states}

To study the effects of quantum fluctuations on the properties of the system, 
we examine the normalized two-body correlation function for the ground states defined as 
\begin{eqnarray}
g^{(2)}(\theta,\theta ')=\frac{\langle\hat{\psi}^{\dagger}(\theta)\hat{\psi}^{\dagger}(\theta ')
\hat{\psi}(\theta ')\hat{\psi}(\theta)\rangle}{\langle\hat{\psi}^{\dagger}(\theta)\hat{\psi}(\theta)\rangle
\langle\hat{\psi}^{\dagger}(\theta ')\hat{\psi}(\theta ')\rangle}\label{cf1}. 
\end{eqnarray}

\begin{figure}[h]
\includegraphics[scale=0.45]{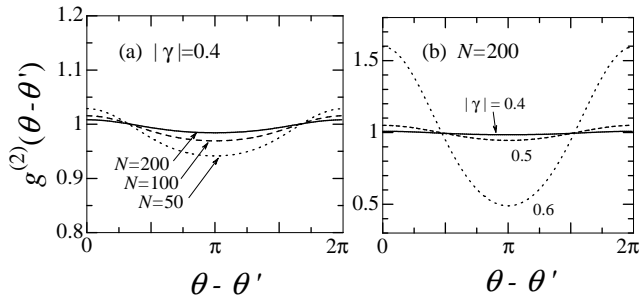}
\caption{Normalized two-body correlation functions $g^{(2)}(\theta-\theta')$ 
(a) for various numbers of particles with a fixed strength of interaction, and 
(b) for various strengths of interactions with a fixed total number of particles. }
\label{corr}
\end{figure}

Figure~\ref{corr}(a) presents $g^{(2)}$ obtained by the exact diagonalization method for several values of $N$ 
for $|\gamma|=0.4$ which is below the critical point. 
The many-body ground state exhibits significant two-body quantum correlations even below the critical 
point, sharply contrasting with the mean-field ground state in which 
$g^{(2)}=\frac{N-1}{N}$ is constant for all $|\gamma|$.
Figure~\ref{corr}(b) shows $g^{(2)}$ for several values of $|\gamma|$ with $N=200$. 
The two-body correlation abruptly increases above the critical point, signaling the formation of a bright soliton.

\begin{figure}
\includegraphics[scale=0.47]{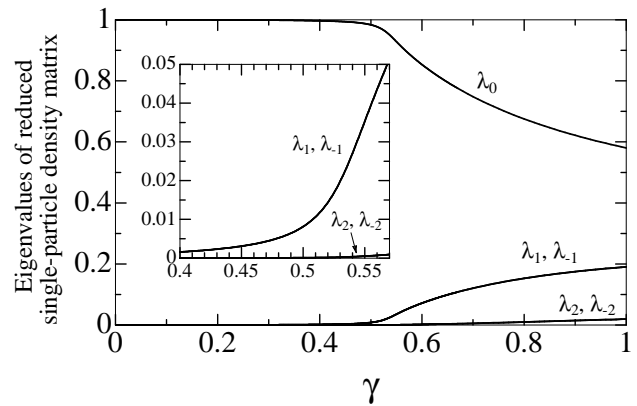}
\caption{Five largest eigenvalues of the reduced single-particle density matrix of the ground state with $N=100$. 
Inset shows the enlarged figure near the critical point. }
\label{eig}
\end{figure}

Figure~\ref{eig} shows the five largest eigenvalues of the reduced single-particle density matrix of the ground state. 
The occupancies of $l=\pm 1$ states increase rapidly above the critical point, indicating the formation of a soliton. 
Remarkably the onset of the increase occurs at values of $|\gamma|$ smaller than $\frac{1}{2}$. 
Although the condensate fraction $\lambda_0$ is unity in the MFT for all $|\gamma|$, 
the many-body calculations reveal that other states are also significantly populated near and beyond the critical point. 
The ground state for $|\gamma|<|\gamma_{\rm cr}|$ corresponds to a single BEC consisting of a 
macroscopic occupation of the zero-momentum state. 
However, when $|\gamma|$ exceeds $\frac{1}{2}$, the condensate fraction 
decreases rapidly and the eigenvalues of higher-momentum states grow 
to make up for the depletion of the condensate. 
The many-body ground state for $|\gamma|>|\gamma_{\rm cr}|$ is thus fragmented~\cite{Noziere}. 
This is a consequence of the exact axisymmetry of the system. 
In the regime $|\gamma| \gg |\gamma_{\rm cr}|$, 
the ground state is expected to change to a non-condensed state 
in which there are no eigenvalues of the reduced density matrix that are of the order of $N$. 
This is consistent with the fact that BEC cannot occur in infinite homogeneous systems.

\begin{figure}
\includegraphics[scale=0.5]{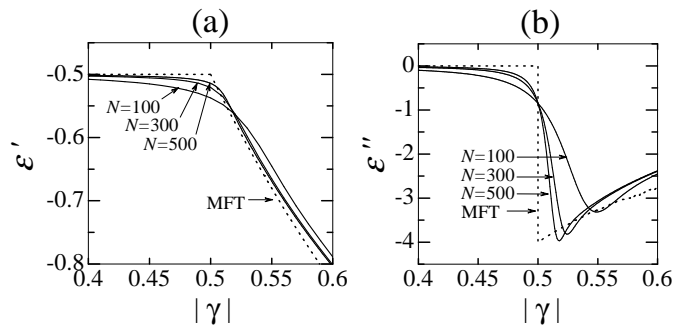}
\caption{(a) First and (b) second derivatives of the many-body ground-state energy for $N=100,300$, and $500$. 
The dotted curves show the corresponding mean-field results.}
\label{derivative}
\end{figure}

The first and the second derivatives of the ground-state energies are shown in Fig.~\ref{derivative}(a) and (b). 
The second derivative of the many-body ground state has no discontinuity unlike the case of the MFT 
(Sec.~\ref{MFcr}), reflecting the fact that strictly speaking no phase transition occurs in finite systems.
The sharp transition that appears in the MFT is replaced by a smooth crossover between the uniform and 
the soliton states. 
The vicinity of $\gamma=-\frac{1}{2}$ may thus be regarded as a crossover regime between a uniform BEC and 
a bright soliton.

\subsection{Low-lying Spectra}\label{excitation}

\begin{figure}[h]
\includegraphics[scale=0.47]{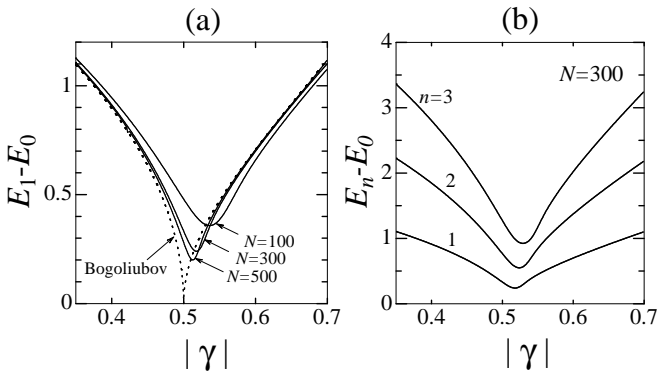}
\caption{(a) First excitation energies $E_1-E_0$ for $N=100,300$, and $500$. 
The dotted curve shows the Bogoliubov first excitation spectrum. 
(b) Three lowest excitation energies for $N=300$.}
\label{EX}
\end{figure}

Figure~\ref{EX}(a) shows the dependence of the lowest excitation energy $E_1-E_0$ on $|\gamma|$ obtained 
by the exact diagonalization method. 
The corresponding Bogoliubov first excitation spectrum, which depends on $N$ 
only through $\gamma$, is superimposed for comparison. 
Because of the conservation of angular momentum, the $l=1$ and $l=-1$ quasiparticles are excited simultaneously. 
The Bogoliubov spectrum in Fig.~\ref{EX}(a) for $|\gamma|\le \frac{1}{2}$ 
is therefore equal to $2\lambda_1$, where $\lambda_1$ is given in Eq.~(\ref{BexU}).

In finite systems we may define the critical point $\gamma_{\rm cr}$ as the one at which 
the lowest excitation energy has a minimum, 
and let $\Delta$ be the minimum of $E_1-E_0$. 
As discussed in Sec.~\ref{MFT}, the mean-field critical point is given by $\gamma_{\rm cr}=-\frac{1}{2}$ and 
the Bogoliubov excitation is gapless at that point. 
The many-body critical point deviates from that of the MFT, the value of $|\gamma_{\rm cr}|$ being 
larger than $\frac{1}{2}$ as shown in Fig.~\ref{EX}. 
In particular, the lowest excitation has an energy gap $\Delta$ at the critical point. 
In the limit of $N\to \infty$, the spectrum gives the same critical exponent 
on both sides of $|\gamma_{\rm cr}|$ as
$E_1-E_0\propto |\gamma-\gamma_{\rm cr}|^{\frac{1}{2}}$ which recovers the result of the MFT.

Figure~\ref{EX}(b) shows the three lowest excitation energies for $N=300$.
We note that $E_n-E_0$ for $n>1$ deviates from $n(E_1-E_0)$ near the critical point, 
which implies that the interaction between the quasiparticles is enhanced. 

\subsection{Energy gap at the critical point}

\begin{figure}
\includegraphics[scale=0.6]{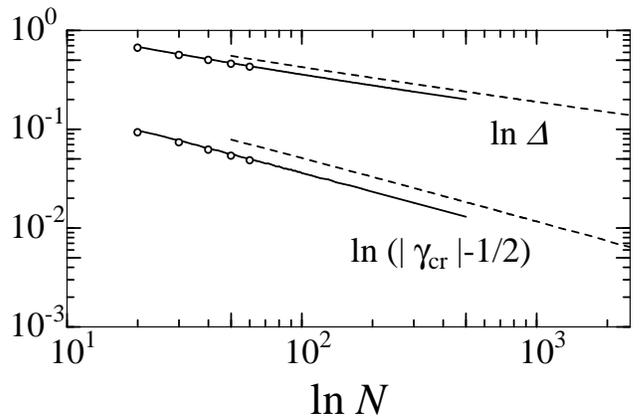}
\caption{Deviation $|\gamma_{\rm cr}|-\frac{1}{2}$ of the many-body critical point from its mean-field value of 
$-\frac{1}{2}$, and the energy gap $\Delta$ obtained by 
the inclusions of the $|l|\le 1$ states (dashed lines), of the $|l|\le 2$ states (solid lines), and 
of the $|l|\le 3$ states (circles).}
\label{gcr}
\end{figure}

As we have seen in Sec.~\ref{excitation}, 
both the critical point $|\gamma_{\rm cr}|$ and the energy gap $\Delta$ that are obtained by 
the exact diagonalization of the Hamiltonian depend on the number of particles $N$. 
The deviation of $\gamma_{\rm cr}$ from $-\frac{1}{2}$ and the value of $\Delta$ are 
shown as functions of $\ln N$ in Fig.~\ref{gcr}. 
The dashed line, the solid line and the circles show the results when the states with $|l|\le 1$, $|l|\le 2$, 
and $|l|\le 3$ are included. 
Figure~\ref{gcr} confirms that (i) the $|l|\le 1$ states give qualitatively correct results, 
(ii) but the inclusion of the $|l|=2$ states in addition to the $|l| \le 1$ states improves the results quantitatively, 
and (iii) the $|l| \ge 3$ states makes few contribution to the results. 

The deviation of $\gamma_{\rm cr}$ from $-\frac{1}{2}$ and the energy gap near the critical point depend on $N$ as
\begin{eqnarray}
|\gamma_{\rm cr}|-\frac{1}{2}\sim N ^{-\frac{2}{3}},\qquad \Delta\sim N^{-\frac{1}{3}}\label{dev}.
\end{eqnarray}
In the large $N$-limit, these values approach those of the MFT, 
that is, $\gamma_{\rm cr}\to -\frac{1}{2}$ and $\Delta\to 0$. 

The power laws in Eq.~(\ref{dev}) can be explained as follows.
In the Bogoliubov theory, we solve Eq.~(\ref{BdG}) assuming that the condensate 
consists of the total number of atoms $N$, and then derive the number of
depleted atoms $N'$ in Eq.~(\ref{depB}).
However, the number of condensed atoms used in Eq.~(\ref{BdG}) should not be $N$ but
$N-N'$.
Therefore, we must solve Eqs.~(\ref{BdG}) and (\ref{depB}) self-consistently when $N'$
is large, {\it i.e.}, near the critical point [Fig.~\ref{BogEX}(b)].
Replacing $\gamma$ in Eq.~(\ref{depB}) with
\begin{eqnarray}
\gamma_0\equiv\frac{(N-N')U}{2\pi}=\gamma\left(1-\frac{N'}{N}\right)\label{difgm},
\end{eqnarray}
we obtain $\frac{N'}{N} \sim N^{-2/3}$ at $\gamma=-\frac{1}{2}$.
From Eq.~(\ref{difgm}), $\gamma_0$ thus behaves as 
\begin{eqnarray}
\gamma_0\sim -\frac{1}{2}+\eta N^{-\frac{2}{3}}, 
\end{eqnarray}
where $\eta$ is a constant. This result reveals that the deviation of the critical point 
from $-\frac{1}{2}$ varies as $N^{-2/3}$, in agreement with 
the numerical result~(\ref{dev}).
Since the energy gap $\Delta$ at the critical point may be regarded as the Bogoliubov excitation 
$\lambda_1\sim \sqrt{\delta}$ as in Eq.~(\ref{BexS}), 
we obtain  $\Delta\sim \sqrt{N^{-2/3}}\sim N^{-1/3}$.
Thus, the behaviors in Eq.~(\ref{dev}) are attributed to depletion of the
condensate due to interactions.
The condition of number conservation in the exact diagonalization method, which treats 
$\hat{N_0}=\hat{c}_0^{\dagger}\hat{c}_0$ as an operator, 
enables us to include the effects of the depletion self-consistently.

\section{Conclusions}

We studied the properties of the ground state and low-lying excitation spectra of attractive 1D BECs 
in a toroidal trap with and beyond the Gross-Pitaevskii MFT. 

In the MFT, the ground states undergo a quantum phase transition between the uniform state and the bright soliton, 
characterized by a jump in the compressibility.
In the Bogoliubov theory, the first excitation spectrum is singular and gapless at the critical point.
Above the critical point, the Goldstone mode appears due to 
the broken symmetry of the soliton state. 
The singularity of the excitation spectrum and the divergence in the number of virtually excited particles 
at the critical point show that the Bogoliubov theory breaks down in the vicinity of the critical point. 

The exact diagonalization of the many-body Hamiltonian removes the singularity at the critical point. 
The prominent effects beyond the MFT are summarized as follows: 
1) The critical point $\gamma_{\rm cr}$ deviates from $-1/2$ and the energy gap $\Delta$ 
emerges in the excitation spectrum. 
2) The singularities in the quantum phase transition are smeared out by quantum fluctuations and 
replaced by a crossover region. 
3) Atoms show two-body correlations for all strengths of interaction. 
4) The condensate fraction begins to decrease from the critical point and the ground state gradually 
changes to a normal state in the strong interaction limit.

Far from the critical point or in the large $N$ limit, 
the MFT offers a good picture of the elementary excitation spectra. 
The deviations of excitation spectra from those of the MFT 
are prominent near the critical point due to the depletion of the condensate. 
The diagonalization method reveals the importance of finite-size effects and the importance of 
the particle-number conservations that lead to qualitative modifications of 
the results of the MFT near the critical point.

\begin{acknowledgments}
This research was supported by a Grant-in-Aid for Scientific Research (Grant No.11216204), by Special 
Coordination Funds for Promoting Science and Technology from the Ministry of Education, 
Culture, Sports, Science and Technology of Japan and by the Toray Science Foundation. 
The matrices were diagonalized by the Lanczos method in the package TITPACK Ver.~2 developed 
by H.~Nishimori. We would like to thank him for this and the useful advice on 
quantum phase transition.
\end{acknowledgments}



\begin{thebibliography}{21}	

\bibitem{ldbec} A.~G\"{o}rlitz, J.~M.~Vogels, A.~E.~Leanhardt, C.~Raman, T.~L.~Gustavson, J.~R.~Abo-Shaeer, 
A.~P.~Chikkatur, S.~Gupta, S.~Inouye, T.~Rosenband, and W.~Ketterle, Phys. Rev. Lett. {\bf 87}, 130402 (2001).

\bibitem {bsoliton1} L.~Khaykovich, F.~Schreck, G.~Ferrari, T.~Bourdel, J.~Cubizolles, L.~D.~Carr, Y.~Castin, C.~Salomon, 
Science {\bf 296}, 1290 (2002).

\bibitem {bsoliton2} K.~E.~Strecker, G.~B.~Partridge, A.~G.~Truscott, and R.~G.~Hulet, Nature {\bf 417}, 150 (2002).

\bibitem{Hoh}P.~C.~Hohenberg, Phys. Rev. {\bf 158}, 383 (1967).

\bibitem{LL} E.~H.~Lieb and W.~Liniger, Phys. Rev. {\bf 130}, 1605 (1963); 
E.~H.~Lieb, Phys. Rev. {\bf 130}, 1616 (1963).

\bibitem{MG} J.~B.~McGuire, J. Math. Phys. {\bf 5}, 622 (1964).

\bibitem{G} M.~D.~Girardeau and E.~M.~Wright, cond-mat/0104585, and references therein.

\bibitem {Carr} L.~D.~Carr, C.~W.~Clark, and W.~P.~Reinhardt, Phys. Rev. A {\bf 62}, 063611 (2000).

\bibitem {ext} J.~G.~Muga and R.~F.~Snider, Phys. Rev. A {\bf 57}, 3317 (1998).

\bibitem{fbMIT} S.~Inouye, M.~R.~Andrews, J.~Stenger, H.~-J.~Miesner, D.~M.~Stamper-Kurn, and W.~Ketterle, 
Nature {\bf 392}, 151 (1998).

\bibitem{fbJILA} S.~L.~Cornish, N.~R.~Claussen, J.~L.~Roberts, E.~A.~Cornell, and C.~E.~Wieman, 
Phys. Rev. Lett. {\bf 85}, 1795 (2000).

\bibitem {toroidal} T.~Kuga, Y.~Torii, N.~Shiokawa, T.~Hirano, Y.~Shimizu, and H.~Sasada, 
Phys. Rev. Lett. {\bf 78}, 4713 (1997).

\bibitem{math} {\it Handbook of mathematical functions}, edited by M.~Abramowitz and I.~A.~Stegun, 
(Dover, New York, 1965).

\bibitem{ZS} V.~E.~Zakharov and A.~B.~Shabat, Sov. Phys. JETP {\bf 34}, 62 (1972).

\bibitem{Noziere} P.~Nozi\`{e}res and D.~Saint James, J. Phys. (Paris) {\bf 43}, 1133 (1982).

\end{thebibliography}
\end{document}